# Revisiting the Polarization Mode Dispersion in Fixed Modulus Model for Random Birefringence

**Vinod Mishra**
*Amvika LLC, 19 Hudson Rd, East Brunswick, NJ 08816*
*vmishra@ieee.org*

**ABSTRACT**

Sometime ago, Galtarossa et al [1] presented some approximate analytical results for the Polarization Mode Dispersion (PMD) assuming a Fixed Modulus Model for the random birefringence. We solve the model exactly and present some new analytical and numerical results.

# 1.INTRODUCTION

The Polarization Mode Dispersion (PMD) [2,5,6] occurs in birefringent optical fibers and acts as an obstacle to higher bit-rates like 40G and 100G. Among many approaches, the spinning and twisting of the optical fibers [1,3,4,7-13] has been found to lessen the PMD. Analytical expressions and numerical results have also been obtained by many authors.

Sometime ago, Galtarossa et al [11] derived expressions for the Differential Group Delay (DGD) of a randomly birefringent fiber in the Fixed Modulus Model (FMM). The DGD has both modulus and the phase. The FMM assumes that the modulus of the birefringence vector is a random variable. They presented analytical results with the following assumptions.

- the spin function is periodic

- the periodicity length (p) of the fiber is much smaller that the Fiber Correlation length ($L_F$) or p<< $L_F$.

Later they also generalized the FMM and presented the Random Modulus Model (RMM), which also includes the randomness in the direction of the birefringence vector. The RMM equations could only be solved numerically.

In the present work the full implications of the FMM has been explored under the following conditions.

- The p<< $L_F$ approximation has been relaxed.
- A nonzero twist has been included.
- A constant spin rate has also been added.

We give the analytical solutions of the exact FMM equations and also present some numerical results showing the effect of different physical conditions.

## 2. THEORETICAL ANALYSIS

### 2.1 The Model

The starting point is the well-known vector equation describing the change in the local PMD vector $\vec{\tau}$ ($\omega$, z) with the angular frequency $\omega$ and distance $z$ along an unspun and untwisted fiber.

$$\frac{\partial \vec{\tau}}{\partial z} = \frac{\partial \vec{\beta}}{\partial \omega} + \vec{\beta} \times \vec{\tau} . \qquad (1)$$

Here $\vec{\beta}(\omega, z)$ is the Local Birefringence Vector (LBV) with initial value $[\beta_l(\omega),0,0]$. Another popular choice for the PMD vector is $\vec{\Omega}$ as used by Galtarossa et al [1].

It is well known that LBV is inherently random because of the imperfections in the fiber. This randomness can be due to statistically fluctuating modulus and phase. In Fixed Modulus Model (FMM), the LBV is modeled as

$$\vec{\beta}(z,\omega) = \beta_l(\omega)(\cos 2\theta(z), \sin 2\theta(z), 0)^T,$$

with superscript T denoting transpose. The modulus is held fixed. The phase is a random quantity obeying the Wiener process described by $\theta(z)$ which obeys the equation

$$\frac{d\theta}{dz} = \sigma \eta(z),$$

with constant $\sigma$ and a white-noise source $\eta(z)$.

Spin and twist can be added to the fiber with Spin Rate function $a(z)$ and externally applied twist rate $\gamma$. After rotating the reference frame to compensate for the rotation induced by the spin, twist and LBV, the set of deterministic equations in Eq. (1) is transformed to a set of stochastic differential equations (SDE)

$$\begin{aligned}
\tau_1' &= \hat{f}\tau_2 + \beta_\omega \\
\tau_2' &= -\hat{f}\tau_1 - \beta_l \tau_3 \\
\tau_3' &= \beta_l \tau_2 + g_\omega \gamma.
\end{aligned} \quad (2)$$

Various symbols occurring in these equations are defined below.

$$\tau_k' = \frac{\partial \tau_k}{\partial z}, k = 1,2,3,$$

$$a'(z) = \frac{\partial a}{\partial z}, \beta_\omega = \frac{\partial \beta_l}{\partial \omega}, g_\omega = \frac{\partial g}{\partial \omega}, \qquad (3)$$

$$\hat{f} = 2a'(z) + (2-g)\gamma + 2\sigma\eta(z).$$

Using Dynkin's formula the above set of equations can be transformed back to a deterministic set again but now they apply for the mean value of the PMD vector components

$$<\tau_1'> = -2\sigma^2 <\tau_1> + f <\tau_2> + \beta_\omega$$

$$<\tau_2'> = -f <\tau_1> - 2\sigma^2 <\tau_2> - \beta_l <\tau_3>$$

$$<\tau_3'> = \beta_l <\tau_2> + g_\omega \gamma,$$

where $f = 2a'(z) + (2-g)\gamma$. When compared with Eq (15) of Galtarossa et al [1] it is seen that a constant due to twist is added to the spin-induced term.

## 2.2 Analytical Results for Spin Rate Function as a constant

We make the approximation that the spin rate function $a'(z) = \partial a / \partial z$ is a constant given by $\alpha_0$. In practical situations this function is more complicated and it is almost impossible to obtain analytical solutions for them.

Under the above approximation, the Eq (1) can be rewritten in the matrix form

$$\frac{d <\vec{t}(s)>}{ds} = \mathbf{A} <\vec{t}(s)> + \mathbf{b}, \qquad (6)$$

in terms of $3 \times 3$ square matrix $\mathbf{A}$ and $3 \times 3$ column matrix $\mathbf{b}$

$$\mathbf{A} = \begin{pmatrix} -c_3 & b_0 & 0 \\ -b_0 & -c_3 & -1 \\ 0 & 1 & 0 \end{pmatrix}, \text{ and } \mathbf{b} = \begin{pmatrix} 1 \\ 0 \\ c_2 \end{pmatrix}. \qquad (5)$$

The various symbols inside the matrices are dimensionless constants and they are explained below.

(i) $q_0 \ (= 2\alpha_0 / \beta_l)$ is the dimensionless spin rate function.

(ii) $c_1 = (2-g)\gamma / \beta_l$ and $c_2 = g_\omega \gamma / \beta_\omega$ are dimensionless constants related to the twist.

(iii) $b_0 = q_0 - c_1$ is a combination useful for calculation.

(iv) $\langle \vec{t} \rangle = (\beta_l / \beta_\omega) \langle \vec{\tau} \rangle$ is dimensionless PMD vector.

(v) $s = \beta_l z$ is dimensionless distance along the fiber.

(vi) $c_3 = 2\sigma^2 / \beta_l$ is dimensionless phase randomness.

The analytical solutions to the set of these three coupled first-order differential equations are given in the Appendix. They take different forms depending upon the values of the above constants.

## 3. NUMERICAL RESULTS

We have chosen the following global input parameters.

| Parameter | Value |
| --- | --- |
| Fiber beat length ($L_B$) | 15 m |
| Initial birefringence vector ($\beta_l = 2\pi/L_B$) | 0.418 /m |
| Twist rotation coefficient (g) | 0.14 |
| Wavelength | 1550 nm |
| Rotational Frequency ($\omega = 2\pi/\lambda$) | 121,609.9355 /ps |
| $\beta_\omega \cong \beta_l/\omega$ | 3.44444 X $10^{-6}$ ps/m |
| $g_\omega = 0.09 \, g/\omega$ | 1.0361 X $10^{-7}$ s |
| Spin rate function linear coefficient ($\alpha_0$) | 0.2/m |
| Twist rate ($\gamma$) | 2 rad/m |

For the zero dimensionless randomness parameter ($c_3 = 2\sigma^2/\beta_l = 0$), the analytical results reduce to the well-known formulas for the DGD. In the presence of nonzero randomness ($c_3 = 2\sigma^2/\beta_l \neq 0$), they depend only on the dimensionless constants $b_0$ and $c_2$.

The physical quantity Differential Group Delay (DGD) is proportional to the length of the PMD vector. It can be expressed in terms of the components $\langle t_i(s) \rangle, i = 1,2,3$ as,

$$\langle \Delta\tau(z) \rangle = \left(\frac{\beta_\omega}{\beta_l}\right)\left[\langle \vec{t}(s) \rangle^2\right]^{1/2} = \left(\frac{\beta_\omega}{\beta_l}\right)\left[\langle t_1(s) \rangle^2 + \langle t_2(s) \rangle^2 + \langle t_3(s) \rangle^2\right]^{1/2}.$$

The no-twist and no-spin DGD depends only on $c_3$ (or randomness in physical terms) and is given as $\{1 - \exp(-c_3 s)\}/c_3$. Then the normalized quantity DGDN is obtained by dividing the DGD by this expression.

$$\text{DGDN}(z) = \frac{\left\{\langle t_1(s) \rangle^2 + \langle t_2(s) \rangle^2 + \langle t_3(s) \rangle^2\right\}^{1/2}}{\left[\left\{\langle t_1(s) \rangle^2 + \langle t_2(s) \rangle^2 + \langle t_3(s) \rangle^2\right\}^{1/2}\right]_{NoTwist, NoSpin}}$$

$$= \frac{\left\{\langle t_1(s) \rangle^2 + \langle t_2(s) \rangle^2 + \langle t_3(s) \rangle^2\right\}^{1/2}}{\{1 - \exp(-c_3 s)\}/c_3}$$

As an example, this expression is plotted for case 2A.1 below using the "Graphmatica"

plotting program.

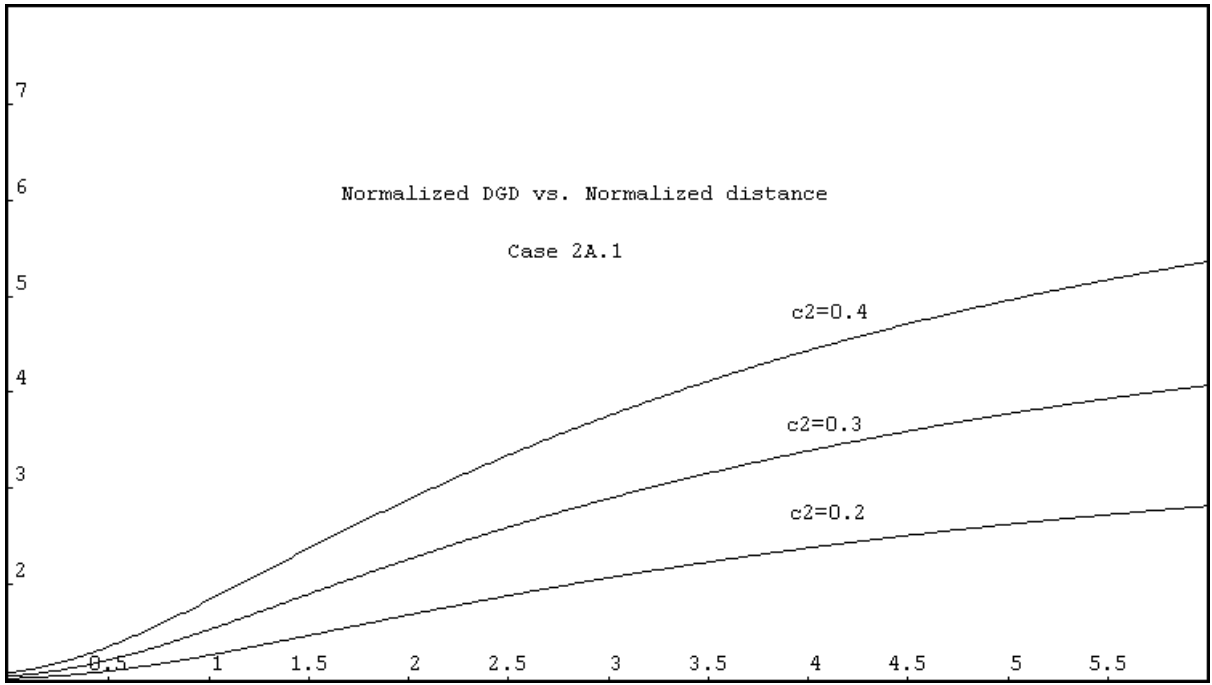

Other cases give similar results.

## 6. CONCLUSIONS

The new expressions are very useful for studying the behavior of DGD for given profile of the spin function. These solutions contain the effects of spin and twist at the same time and interestingly in those cases, the solutions only depend on their combinations. Further studies should make it possible to discover new regimes of the spin and twist values where the DGD may have very low values. This will be reported in future.

## 7. ACKNOWLEDGEMENT

I thank Nick Frigo (AT&T Labs) for getting me interested in this topic.

# APPENDIX

The equations obtained after expanding Eq. (6) are

$$\frac{d}{ds}<t_1(s)> = -c_3 <t_1(s)> + b_0 <t_2(s)> +1 \quad \text{(A.1a)}$$

$$\frac{d}{ds}<t_2(s)> = -b_0 <t_1(s)> - c_3 <t_2(s)> - <t_3(s)> \quad \text{(A.1b)}$$

$$\frac{d}{ds}<t_3(s)> = <t_2(s)> + c_2, \quad \text{(A.1c)}$$

One can assume that the initial DGD vector is zero without any loss of generality so the boundary conditions are given by $<t_i(0)> = 0, i = 1,2,3$. Defining the Laplace Transform pairs as,

$$<\tilde{t}_i(u)> = \int_0^\infty \exp(-su) <t_i(s)> ds, \quad \text{(A.2)}$$

and using it on (A.1) gives,

$$(u+c_3)<\tilde{t}_1(u)> = b_0 <\tilde{t}_2(u)> + (1/u) \quad \text{(A.3a)}$$

$$(u+c_3)<\tilde{t}_2(u)> = -b_0 <\tilde{t}_1(u)> - <\tilde{t}_3(u)> \quad \text{(A.3b)}$$

$$u<\tilde{t}_3(u)> = <\tilde{t}_2(u)> + (c_2/u). \quad \text{(A.3c)}$$

In presence of randomness $(c_3 = 2\sigma^2/\beta_l \neq 0)$, the analytical results depend only on the two dimensionless constants $b_0$, and $c_2$. Solving the above algebraic equations we get,

$$<\tilde{t}_i(u)> = (\bar{a}_i + \bar{b}_i u + \bar{c}_i u^2)/D(u), i = 1,2,3.$$

with

$$\bar{a}_1 = 1 - b_0 c_2, \bar{b}_1 = c_3, \bar{c}_1 = 1$$

$$\bar{a}_2 = -c_2 c_3, \bar{b}_2 = -(b_0 + c_2), \bar{c}_2 = 0$$

$$\bar{a}_3 = c_2(b_0^2 + c_3^2) - b_0, \bar{b}_3 = 2c_2c_3, \bar{c}_3 = c_2$$

And the denominator is

$$D(u) = u^3 + 2c_3u^2 + (b_1^2 + c_3^2)u + c_3 = (u+v)(u+v')(u+w)$$

with $b_1^2 = 1 + b_0^2 = 1 + (q_0 - c_1)^2$. The denominator is a cubic and depending on the value of the coefficients, it has different kinds of roots. We define below some cubic-related auxiliary symbols.

$$y_N = c_3(1 - \tfrac{2}{3}b_1^2 - \tfrac{2}{27}c_3^2)$$

$$\delta = \tfrac{1}{3}(c_3^2 - 3b_1^2)^{1/2},$$

$$h = 2\delta^3$$

Then

$$y_N^2 - h^2 = \frac{1}{27}\left[4b_1^2(b_1^2 + c_3^2)^2 + 27c_3^2 - 4c_3^2(c_3^2 + 9b_1^2)\right]$$

is the expression distinguishing different roots. The solutions depending on these different situations follow.

**Case 1:** $y_N^2 < h^2$, Three distinct real roots ($v \neq v' \neq w$)

Define $\cos 3\theta = -y_N/h$. The roots are given as

$$v = \tfrac{2}{3}c_3 - 2\delta\cos\theta$$

$$v' = \tfrac{2}{3}c_3 - 2\delta\cos(\theta + \tfrac{2\pi}{3})$$

$$w = \tfrac{2}{3}c_3 - 2\delta\cos(\theta + \tfrac{4\pi}{3})$$

After some algebra, the solutions are given as follow ($i = 1,2,3$).

$$<t_i(s)> = \frac{\bar{a}_i}{vv'w} - \frac{1}{v'-v}\left(\frac{\bar{a}_i}{vw} + \frac{(\bar{a}_i/w) - \bar{b}_i + \bar{c}_i v}{w-v}\right)\exp(-vs)$$

$$+ \frac{1}{v'-v}\left(\frac{\bar{a}_i}{v'w} + \frac{(\bar{a}_i/w) - \bar{b}_i + \bar{c}_i v'}{w-v'}\right)\exp(-v's) - \frac{(\bar{a}_i/w) - \bar{b}_i + \bar{c}_i w}{(w-v)(w-v')}\exp(-ws)$$

**Case 2**: $y_N^2 > h^2$ One real root ($w$) and one complex-conjugate pair of root ($v, v^*$).

$$w = \tfrac{2}{3}c_3 - (p+q)$$

$$v = v_r - iv_i$$

$$v^* = v_r + iv_i$$

where

$$v_r = \tfrac{2}{3}c_3 + \tfrac{1}{2}(p+q)$$

$$v_i = \tfrac{\sqrt{3}}{2}|p-q|$$

$$p + q = -(y_N + \sqrt{y_N^2 - h^2})^{1/3} - (y_N - \sqrt{y_N^2 - h^2})^{1/3}$$

$$|p - q| = (y_N + \sqrt{y_N^2 - h^2})^{1/3} - (y_N - \sqrt{y_N^2 - h^2})^{1/3}$$

After some algebra, the solutions are given as follow ($i = 1,2,3$).

$$<t_i(s)> = \frac{\bar{a}_i/w}{v_r^2 + v_i^2}\left[1 - \left(\frac{v_r}{v_i}\sin v_i s + \cos v_i s\right)\exp(-v_r s)\right]$$

$$+ \frac{\bar{a}_i/w - \bar{b}_i - \bar{c}_i v_r}{(w-v_r)^2 + v_i^2}\left(\frac{w-v_r}{v_i}\sin v_i s + \cos v_i s\right)\exp(-v_r s)$$

$$+ \frac{\bar{c}_i}{(w-v_r)^2 + v_i^2}[(w-v_r)\cos v_i s + v_i \sin v_i s]\exp(-v_r s) - \frac{(\bar{a}_i/w) - \bar{b}_i + \bar{c}_i w}{(w-v_r)^2 + v_i^2}\exp(-ws)$$

The above expression can also be written in a different but useful way.

$$<t_i(s)> = \frac{\bar{a}_i/w}{v_r^2+v_i^2} + \left[-\left(\frac{\bar{a}_i}{w}\right)\frac{v_r}{v_r^2+v_i^2} + \frac{(\bar{a}_i/w-\bar{b}_i-\bar{c}_i v_r)(w-v_r)+\bar{c}_i v_i^2}{(w-v_r)^2+v_i^2}\right]\frac{\sin v_i s}{v_i}\exp(-v_r s)$$

$$+\left[-\left(\frac{\bar{a}_i}{w}\right)\frac{1}{v_r^2+v_i^2} + \frac{\bar{a}_i/w-\bar{b}_i+\bar{c}_i(w-2v_r)}{(w-v_r)^2+v_i^2}\right]\cos v_i s\exp(-v_r s) - \frac{(\bar{a}_i/w)-\bar{b}_i+\bar{c}_i w}{(w-v_r)^2+v_i^2}\exp(-ws)$$

**Case 3a**: $y_N^2 = h^2, \delta \neq 0$, One real double root ($v=v'$), one single root ($w$)

$$v = v' = \tfrac{2}{3}c_3 - \delta$$

$$w = \tfrac{2}{3}c_3 + 2\delta$$

After some algebra, the solutions are given as follow ($i=1,2,3$).

$$<t_i(s)> = \frac{\bar{a}_i}{v^2 w} - \left[\frac{(\bar{a}_i/v)\{(w/v)-2\}+\bar{b}_i-\bar{c}_i w}{(w-v)^2} + \frac{(\bar{a}_i/v)-\bar{b}_i+\bar{c}_i v}{w-v}s\right]\exp(-vs)$$

$$-\frac{(\bar{a}_i/w)-\bar{b}_i+\bar{c}_i w}{(w-v)^2}\exp(-ws)$$

**Case 3b**: $y_N^2 = h^2, y_N = 0, \delta = 0$, One real triple root ($v=v'=w=\tfrac{2}{3}c_3$)

After some algebra, the solutions are given as follow ($i=1,2,3$).

$$<t_i(s)> = \frac{\bar{a}_i}{v^3} - \left[\frac{\bar{a}_i}{v^3} + \left(\frac{\bar{a}_i}{v^2}-\bar{c}_i\right)s + \left(\frac{\bar{a}_i}{v}-\bar{b}_i+\bar{c}_i v\right)\left(\frac{s^2}{2}\right)\right]\exp(-vs)$$

For this special case, $c_3 = \tfrac{3}{4}\sqrt{6}, v=v'=w=\tfrac{1}{2}\sqrt{6}$ so the above expression simplifies.

$$<t_i(s)> = \frac{2\sqrt{6}}{9}\left[\bar{a}_i - \left\{\bar{a}_i + \frac{\sqrt{6}}{2}\left(\bar{a}_i - \frac{3}{2}\bar{c}_i\right)s + \frac{3}{2}\left(\bar{a}_i - \frac{\sqrt{6}}{2}\bar{b}_i + \frac{3}{2}\bar{c}_i\right)\left(\frac{s^2}{2}\right)\right\}\exp\left(-\frac{\sqrt{6}}{2}s\right)\right]$$

All the expressions given above have proper limits in absence of randomness ($c_3 = 2\sigma^2/\beta_l = 0$) and reduce to the well-known result.

$$<t_i(s)> = \frac{\bar{a}_i}{b_1^2}s + \left(\bar{c}_i - \frac{\bar{a}_i}{b_1^2}\right)\frac{\sin b_1 s}{b_1} + \frac{\bar{b}_i}{b_1^2}(1-\cos b_1 s)$$

Here $b_1^2 = 1 + b_0^2$.